\documentclass[12pt]{article}
\usepackage{fullpage,epsfig,fancyheadings}
\usepackage[psamsfonts]{euscript}
\usepackage{epic}
\usepackage{amstex}

\newcommand{\beq}{\begin{equation}}
\newcommand{\eeq}{\end{equation}}

\begin{document}
\vspace*{-.6in}
\thispagestyle{empty}
\begin{flushright}
CALT-68-2108\\
DOE RESEARCH AND\\
DEVELOPMENT REPORT
\end{flushright}
\baselineskip = 20pt

\vspace{.5in}
{\Large
\begin{center}
Comparison of Lattice and Dual QCD Results for Heavy Quark Potentials
\end{center}}
\vspace{.4in}

\begin{center}
M. Baker\\
\emph{University of Washington, Seattle, Washington  98105}
\end{center}

\begin{center}
James S. Ball\\
\emph{University of Utah, Salt Lake City, Utah}
\end{center}

\begin{center}
F. Zachariasen\\
\emph{California Institute of Technology, Pasadena, CA  91125}
\end{center}
\vspace{1in}

\begin{center}
\textbf{Abstract}
\end{center}
\begin{quotation}
\noindent  Lattice results and Dual QCD results for all heavy quark potentials
through order (quark mass)$^{-2}$ are exhibited and compared.  The agreement on
the whole is quite good, confirming the validity of Dual QCD.
\end{quotation}
\vfil

\newpage

\pagenumbering{arabic} 


Bali, et al.,~\cite{1} have recently calculated from lattice theory all of the
heavy quark potentials --- the central potential, all spin dependent
potentials, and all velocity dependent potentials, through order velocity
squared, or, equivalently, through order (quark mass)$^{-2}$.  We have
previously computed all of these same potentials from the Dual Superconducting
model of QCD, (i.e.) Dual QCD~\cite{2,3}.  Our purpose in this note is to
compare this new lattice data with Dual QCD predictions.

The definitions of the potentials by Bali, et al.,~\cite{1} are the same as in
Dual QCD, except for those proportional to velocity squared.  (Bali, et al.,
include in their calculation some numbers called $c_2, c_3, c_4$ etc. which
represent ratios of the running coupling $\alpha_s$ at various energies.  We
have set all these ratios equal to one because in dual QCD the coupling
constant, in the classical approximation used to derive the potentials, does
not run.)  The comparison of the potentials is given in Table 1.

\begin{center}
{\bf Table 1}
\end{center}
\begin{table}[h]
\begin{center}
\begin{tabular}{l|l}
\qquad \qquad \qquad{\bf Bali, et al.} & \qquad \qquad{\bf Dual QCD}\\
$V_0 + {1\over 8} \left({1\over m_1^2} + {1\over m_2^2}\right) (\nabla^2 V_0 +
\nabla^2 V_a^E - \nabla^2 V_a^B)$ & $V_0 + {1\over 8} \left({1\over m_1^2} +
{1\over m_2^2}\right) (\nabla^2 V_0 + \nabla^2 V_a)$\\
$V_1'$ & $V_1'$\\
$V_2'$ & $V_2'$\\
$V_3 $ & $V_3$\\
$V_4 $ & $V_4$\\
$V_b $ & ${1\over 3} (-V_+ + V_- + {1\over 2} V_{\parallel} - {1\over 2}
V_L)$\\
$V_c$ & ${1\over 2} (- V_+ + V_- - V_{\parallel} + V_L)$\\
$V_d$ & ${1\over 6} (V_+ + V_- + {1\over 2} V_{\parallel} + {1\over 2} V_L)$\\
$V_e$ & ${1\over 4} (V_+ + V_- - V_{\parallel} - V_L)$\\
\end{tabular}
\end{center}
\end{table}

In dual QCD, the potential $V_a$ can also be broken up into an electric and a
magnetic part:
\begin{equation}\label{1}
V_a = V_a^E - V_a^B
\end{equation}
where~\cite{3}
\begin{equation}\label{2}
\nabla^2 V_a^E = - \nabla^2 V_0^{NP}
\end{equation}
and
\begin{equation}\label{3}
\nabla^2 V_a^B = - {4\over 3} e^2 \vec\nabla \cdot \vec\nabla' G^{NP} (\vec x,
\vec x')\big|_{\vec x = \vec x' = z_j}.
\end{equation}
(Here the superscript $NP$ stands for nonperturbative.)  The dual QCD result
for $V_a^B$ is weakly singular and requires a cutoff~\cite{4}.  The same result obtains for the lattice calculation of $V_a^B$~\cite{6}.  The spin-spin potential $V_{4}$ has a
delta function term and the term $\nabla^2 (V_0 + V^E_a)$ in
dual QCD is simply proportional to a delta function at the origin~\cite{3},
though these naturally do not show up cleanly in the lattice calculation.  All
of the remaining potentials are finite and well behaved in both approaches.

The comparison of the two sets of results are shown in Figures 1 through
10.  Fig. 1 shows the lattice and the dual QCD calculations of the central potential
$V_0 (R)$.  The units are GeV and Fermis.  The dual QCD parameters given in
Reference (2) have been changed to produce a best  fit to the lattice $V_{0}$
for $\beta=6.2$.  The new parameters
are $\alpha_s = .2048$ and the string tension $\sigma = .2384 GeV^2$.
These changes  significantly worsen the
fits for the $c\bar c$ and $b\bar b$ spectra given in Reference (2).  The 
resulting effective $\chi^{2}$ is 11.4, about 6 times that of our earlier fit.
The average error increases from 13 MeV to 29 MeV.  While our method of calculation differs considerably from that used by Bali, et. al. the quality of our
fit described here is comparable to theirs.

The next figure shows the comparison of the quantity $\nabla^2 V_a^E$.  The
agreement, evidently, is not bad.  We recall, however, as mentioned before,
that in dual QCD $\nabla^2 (V_0 + V_a^E)$ is simply a delta function.  This
result does not hold on the lattice, so some discrepancy in $\nabla^2 V_a^E$,
especially at small $R$, is not surprising.

There is no figure for $\nabla^2 V_a^B$, because, also as mentioned above, in
both dual QCD and on the lattice this quantity is weakly divergent but is not very sensitive to 
the required  cutoff.  A detailed analysis and comparison of $\nabla^2 V_a$ in
dual QCD and on the lattice is given in Reference (5).

The remaining Figures (3 through 10) show the parameter free Dual QCD predictions and lattice data 
for the rest of the potentials, namely $V_1', V_2', V_3, V_4, V_b, V_c, V_d$
and $V_e$.  All of these agree remarkably well (within the lattice calculation
uncertainties), with a few relatively minor exceptions.  The short distance behavior of $- \nabla^2 V_a^E$ and $V_3$ is above the lattice data.  In this domain radiative corrections giving rise to running coupling constants and asymptotic freedom become important, and those are not included in dual QCD.  Also, at small
$R$, Figure 5  shows the dual QCD spin-spin potential to be well above the lattice
points.  To understand the possible origin of this difference, consider 
the  interaction of a point magnetic dipole with a sphere of constant magnetization in which dipole and magnetization directions are determined by the two spin directions.  For the dipole outside of the sphere the interaction potential 
is of the form of $V_{3}$ and  produces  the usual perturbative QCD result.  
For the dipole inside the sphere the interaction is a constant and has the spin
dependence of $V_{4}$.  If one takes the radius of the sphere to zero holding
its magnetic moment constant this potential becomes a delta function at the
origin.  Because of the fact that the finite lattice size represents a 
granularity in space, one might expect a modification of the small R
behavior of both of these potentials in a lattice calculation. 

Figures 9 and 10 for $V_d$ and $V_e$ show that 
the lattice results are consistent with zero for these two potentials.  The
dual QCD results are also nearly flat and very small, so we agree with what one
gets on the lattice.

To summarize, the lattice data for the central potential $V_0(R)$ was used to determine the parameters $\alpha_s$ and $\sigma$ of dual QCD.  The resulting fit is shown in Fig. 1.  All the remaining potentials are then uniquely predicted (Figs. 2-10).  Overall the agreement of these dual QCD predictions with the lattice data is remarkably good, and we feel that it provides evidence for the validity of the dual picture of long distance Yang--Mills theory. 

We would like to thank Gunnar Bali for making their lattice results available to us in the numerical form necessary for the detailed fits and comparison.  
\newpage

\begin{center}
{\bf Figure Captions}
\end{center}

\begin{description}
\item {\bf Fig. 1}  Comparison of the dual QCD central potential (solid line)
with the lattice central potential (points)for $\beta=6.2$.  The dual QCD parameters are
$\alpha_s = .2048$, and $\sigma = - .2384 (GeV)^2$ .  The
lattice string tension, in contrast, is $\sigma = .2190 (GeV^2)$.

\item {\bf Fig. 2}  A similar comparison (using the same parameters) for the
quantity $\nabla^2 V_a^E$.

\item {\bf Fig. 3}  The same for $-V'_1$.  (Note the minus sign.)

\item {\bf Fig. 4}  The same for $V'_2$.

\item {\bf Fig. 5}  The same for $V_3$.

\item {\bf Fig. 6}  The lattice result for the lowest value of $R$ has been omitted as it contains the delta function contribution to the potential.  Our curve is the contribution of dual QCD other than the delta function. 

\item {\bf Fig. 7}  And for $V_b$.

\item {\bf Fig. 8}  And for $V_c$.

\item {\bf Fig. 9}  And for $V_d$.

\item {\bf Fig. 10}  And finally the comparison for $V_e$.
\end{description}

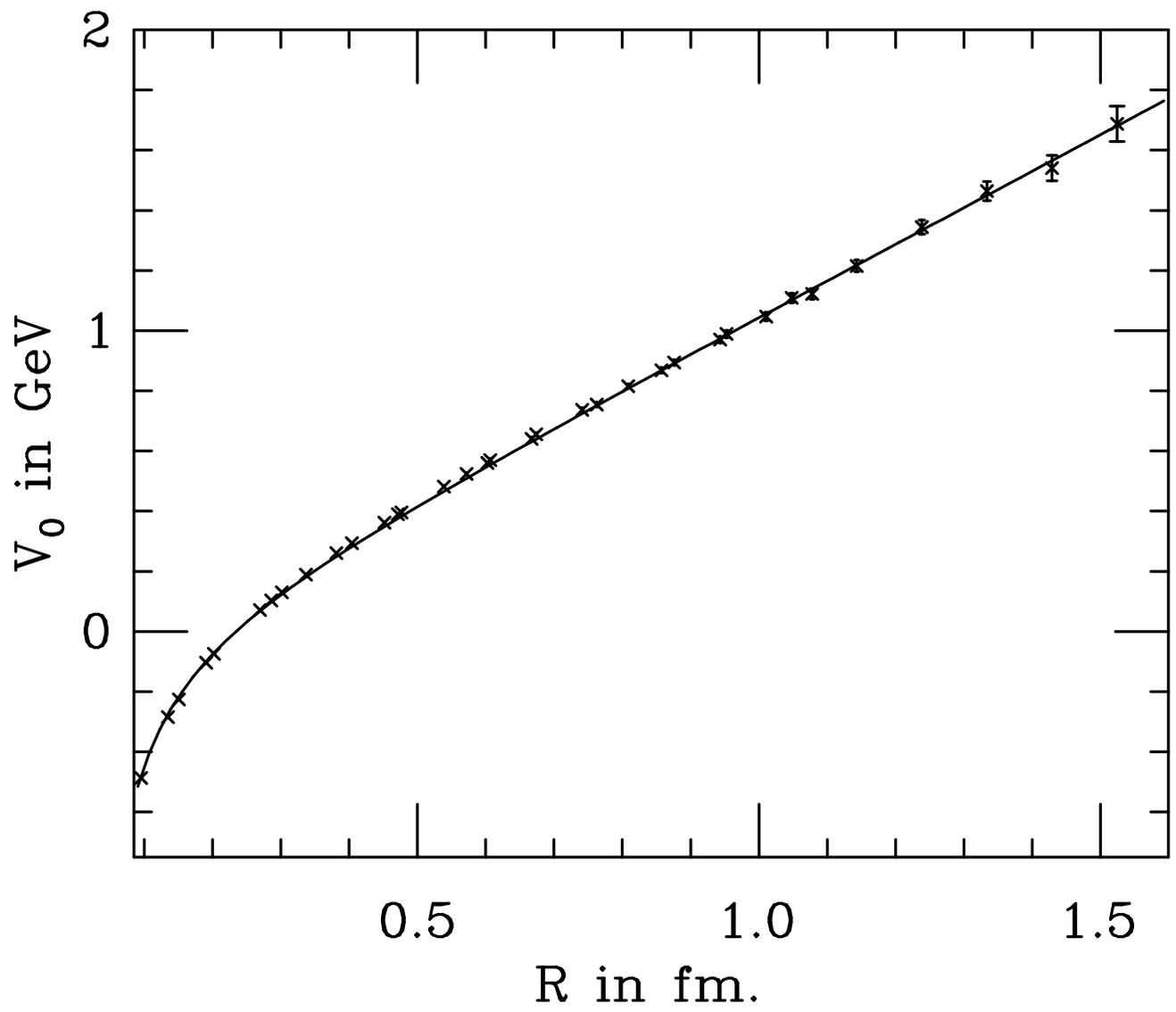
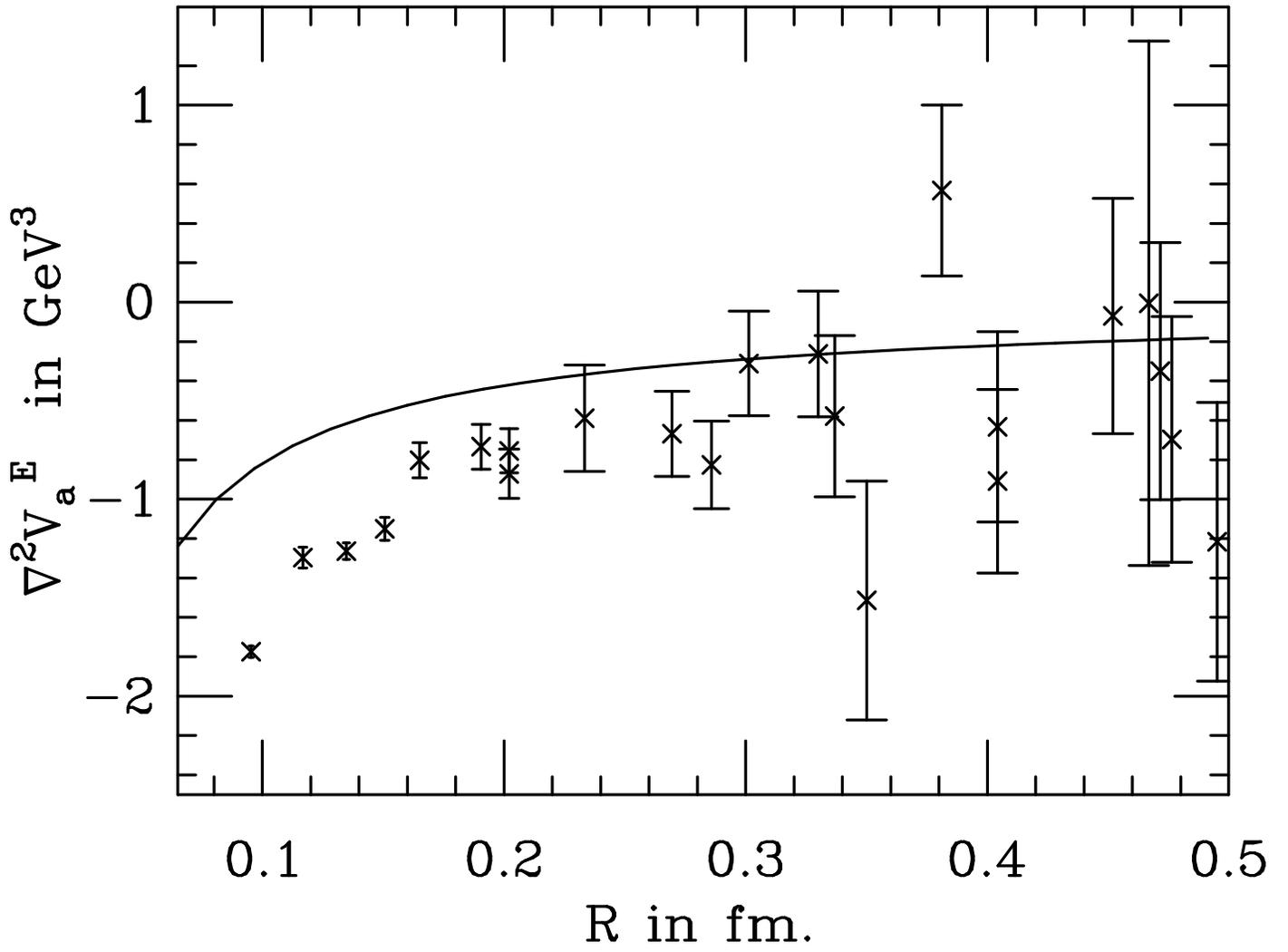
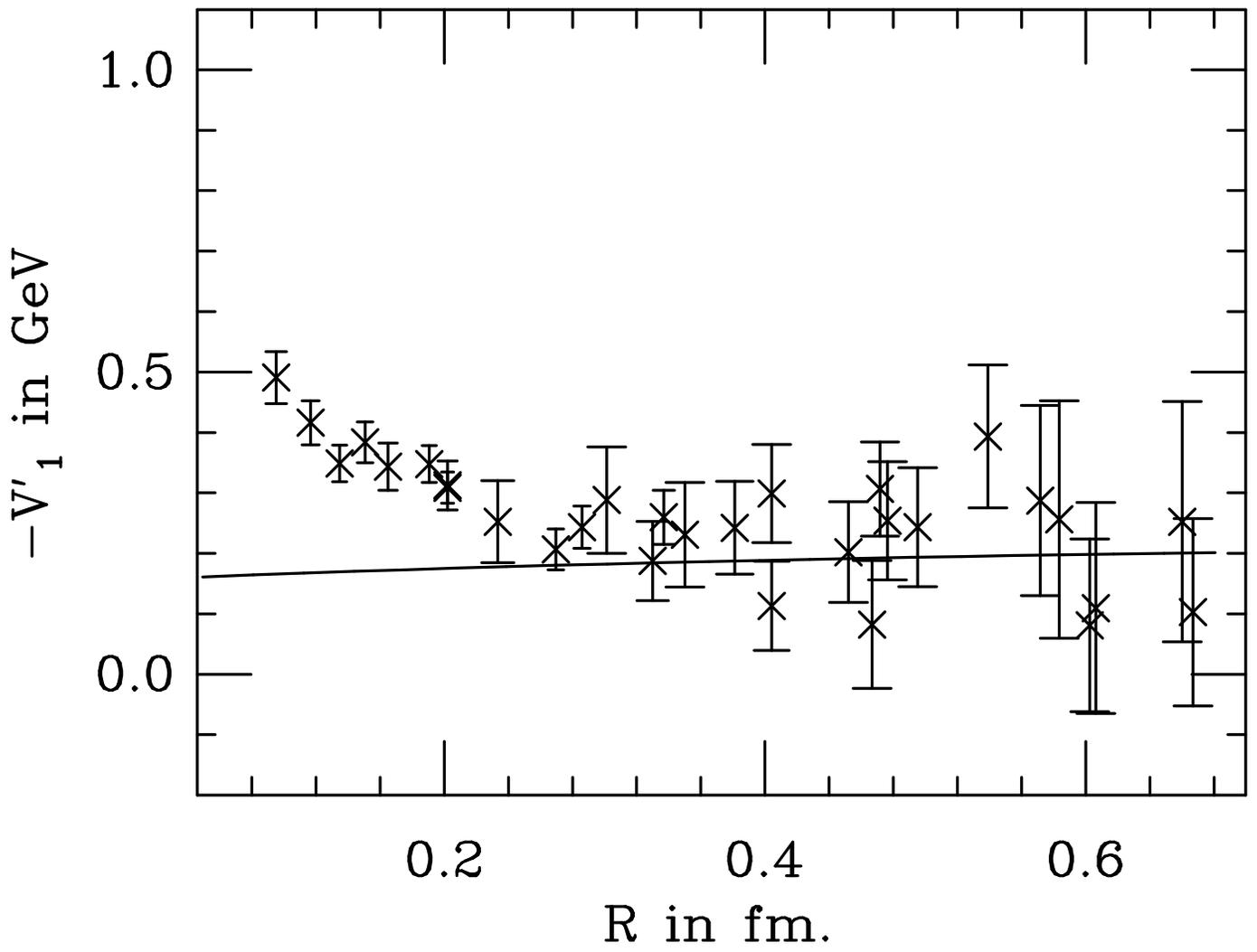
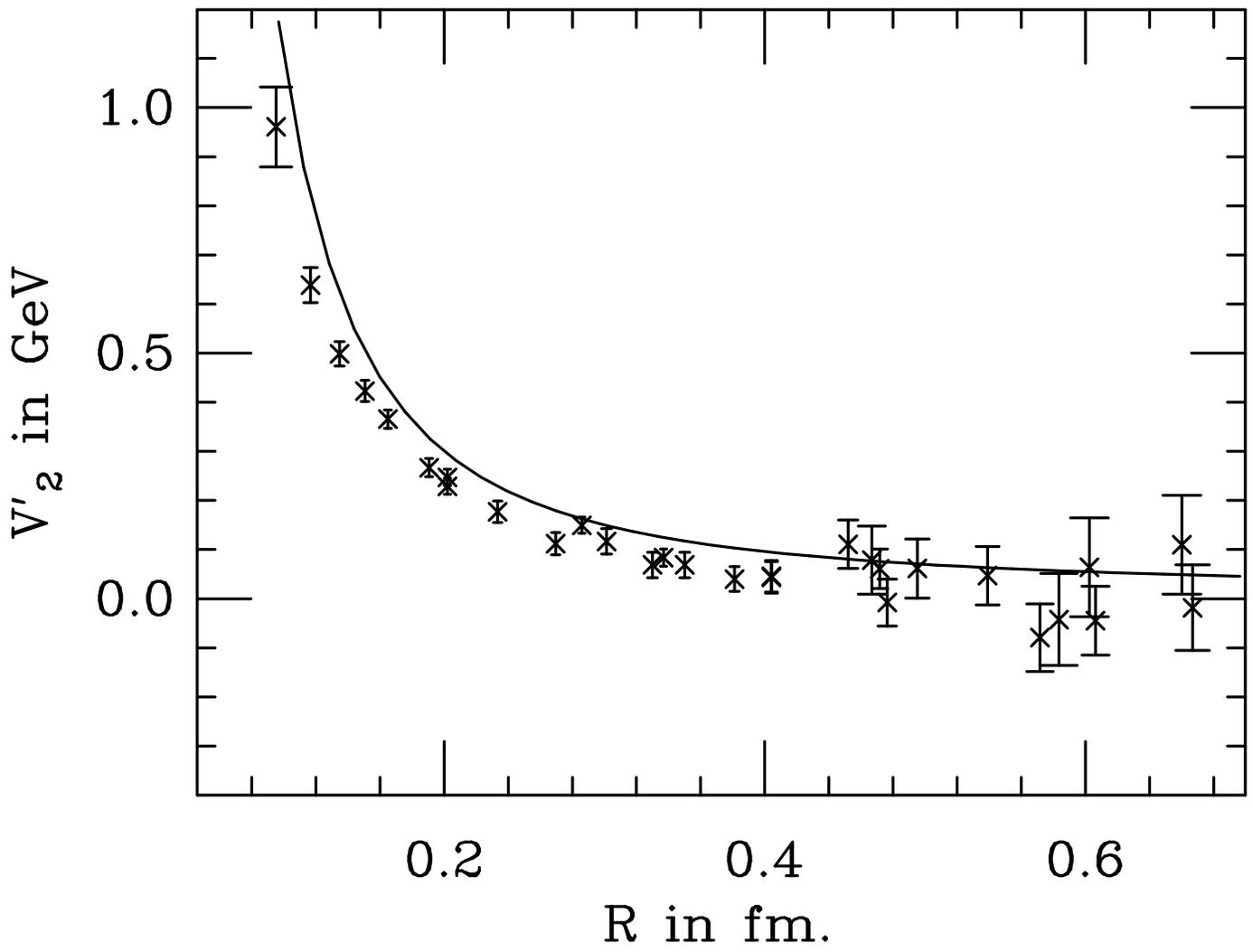
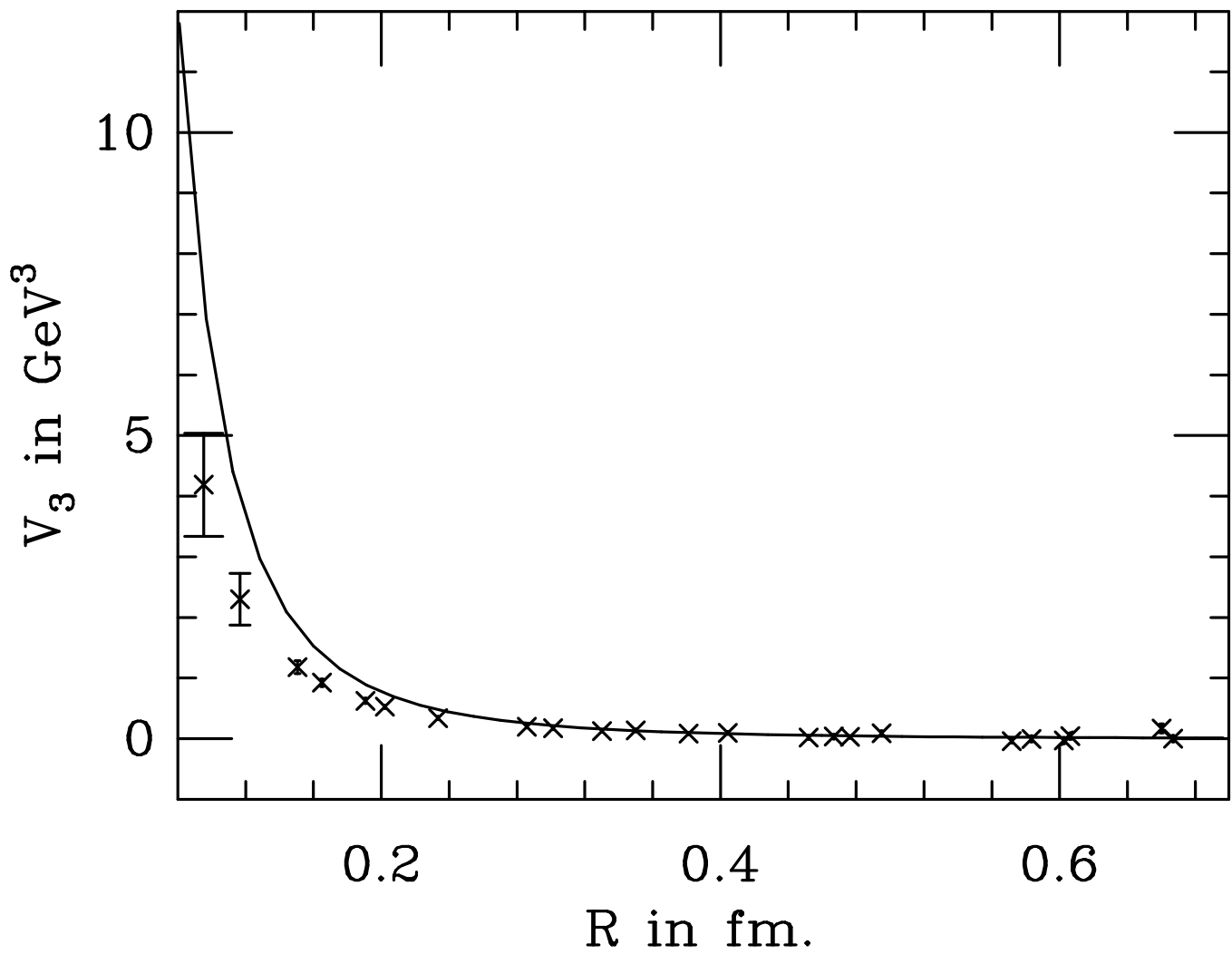
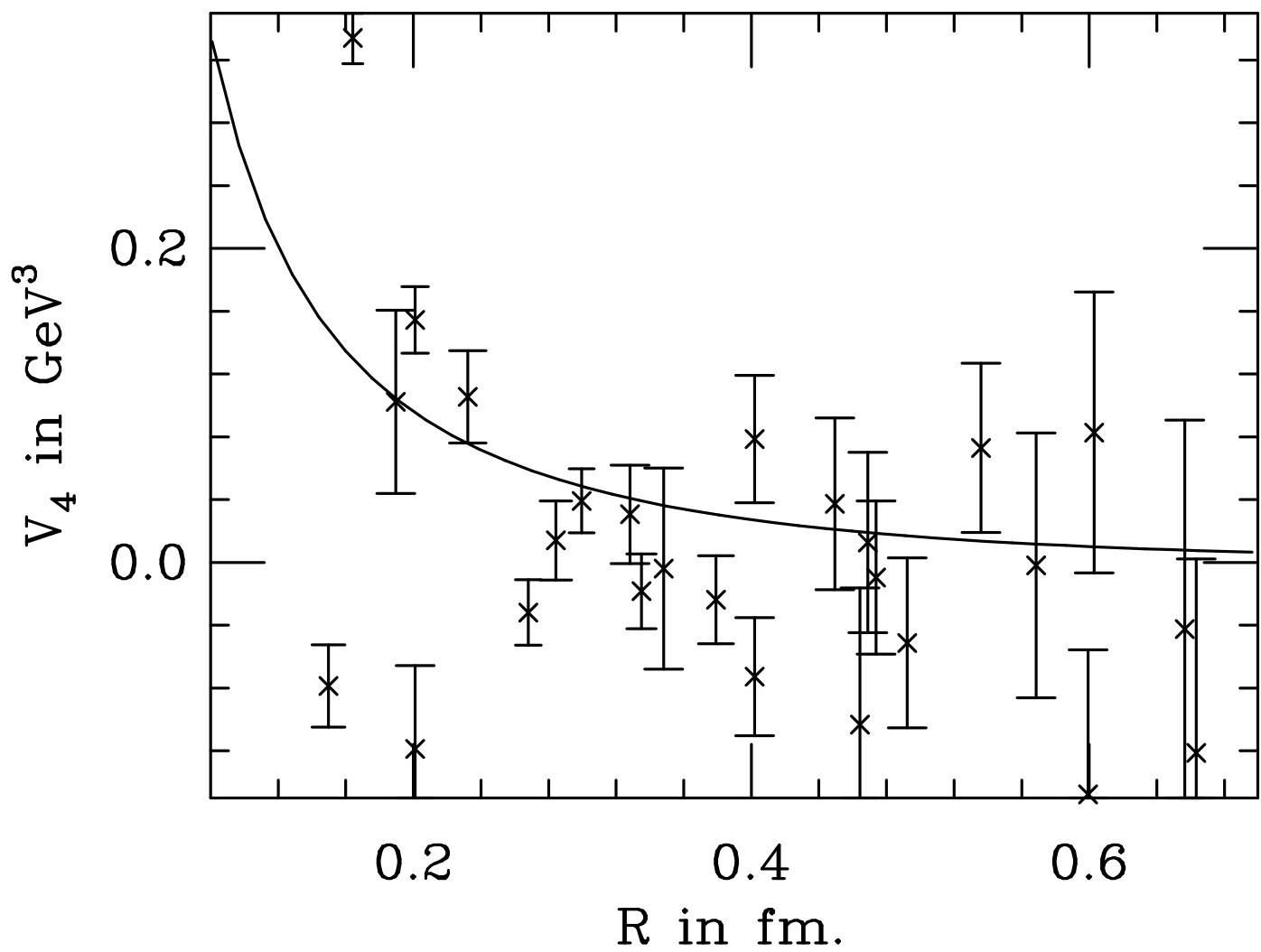
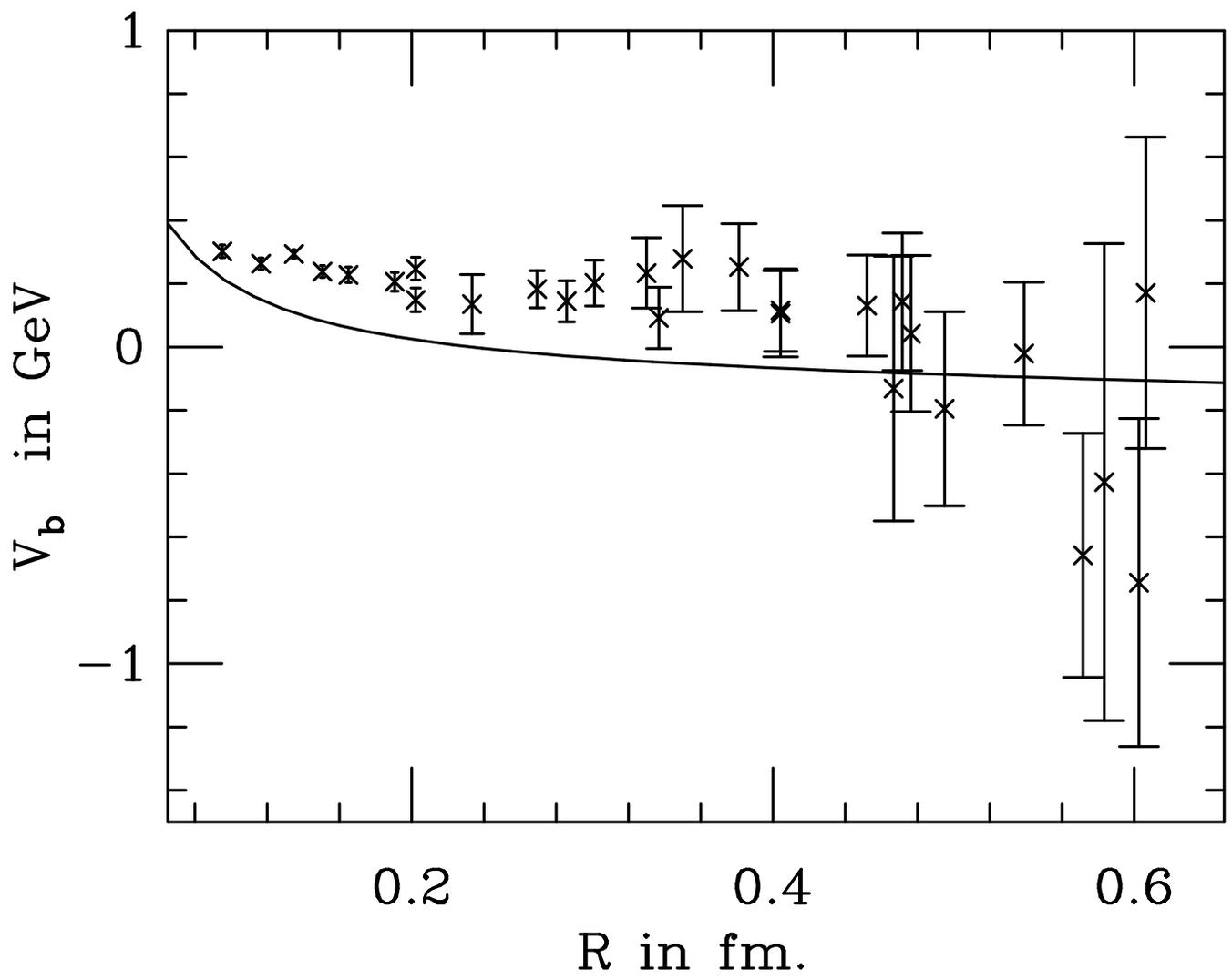
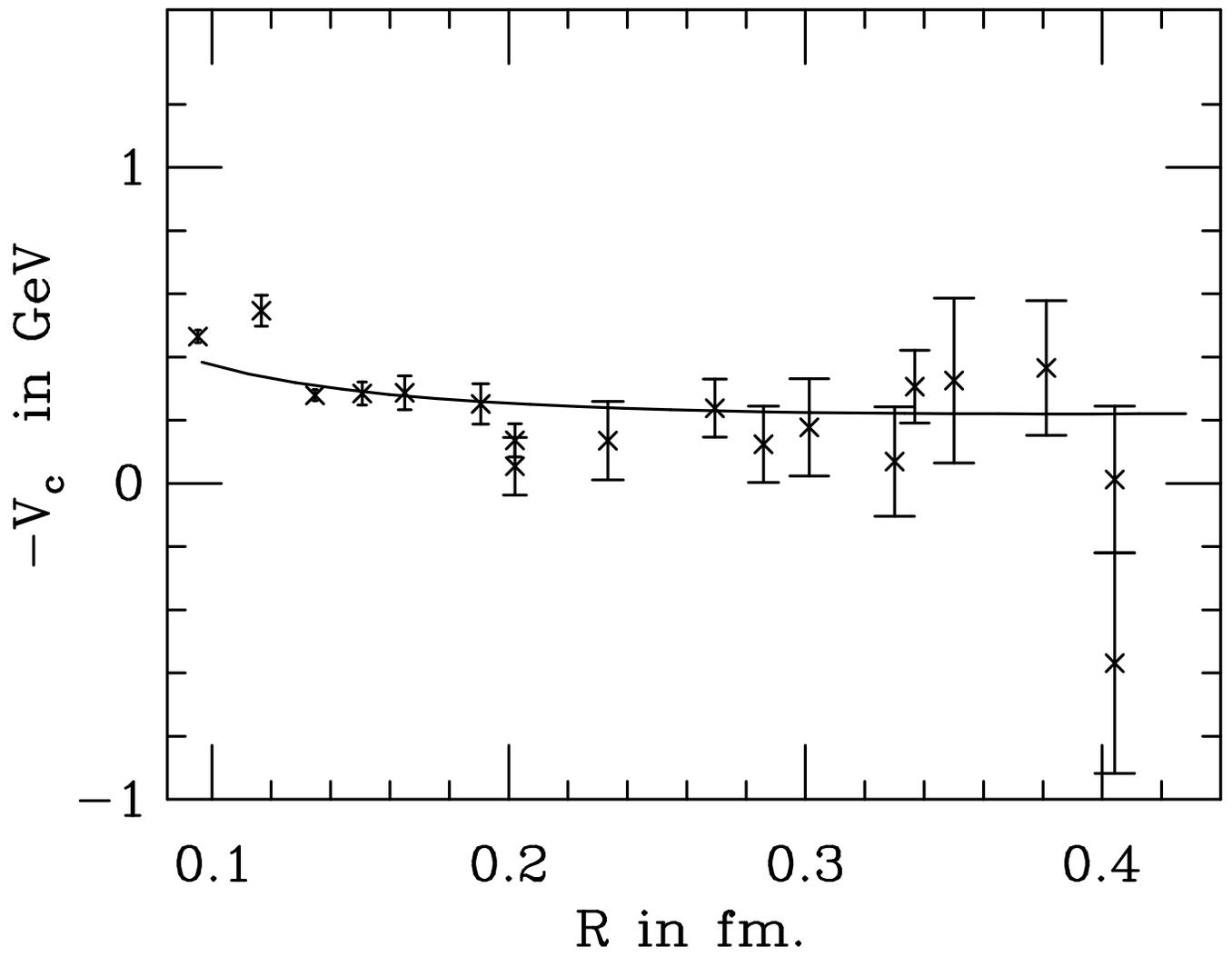
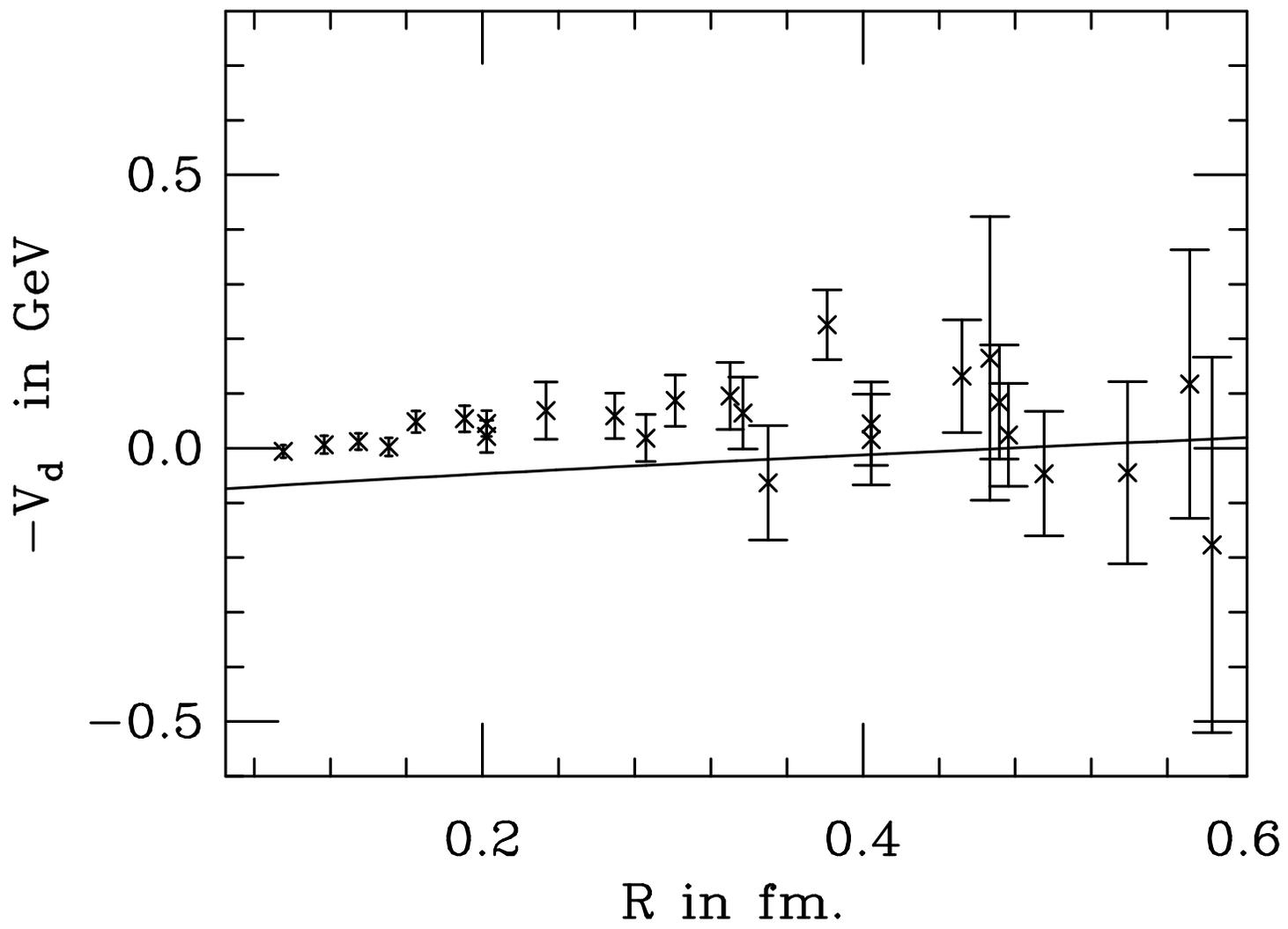
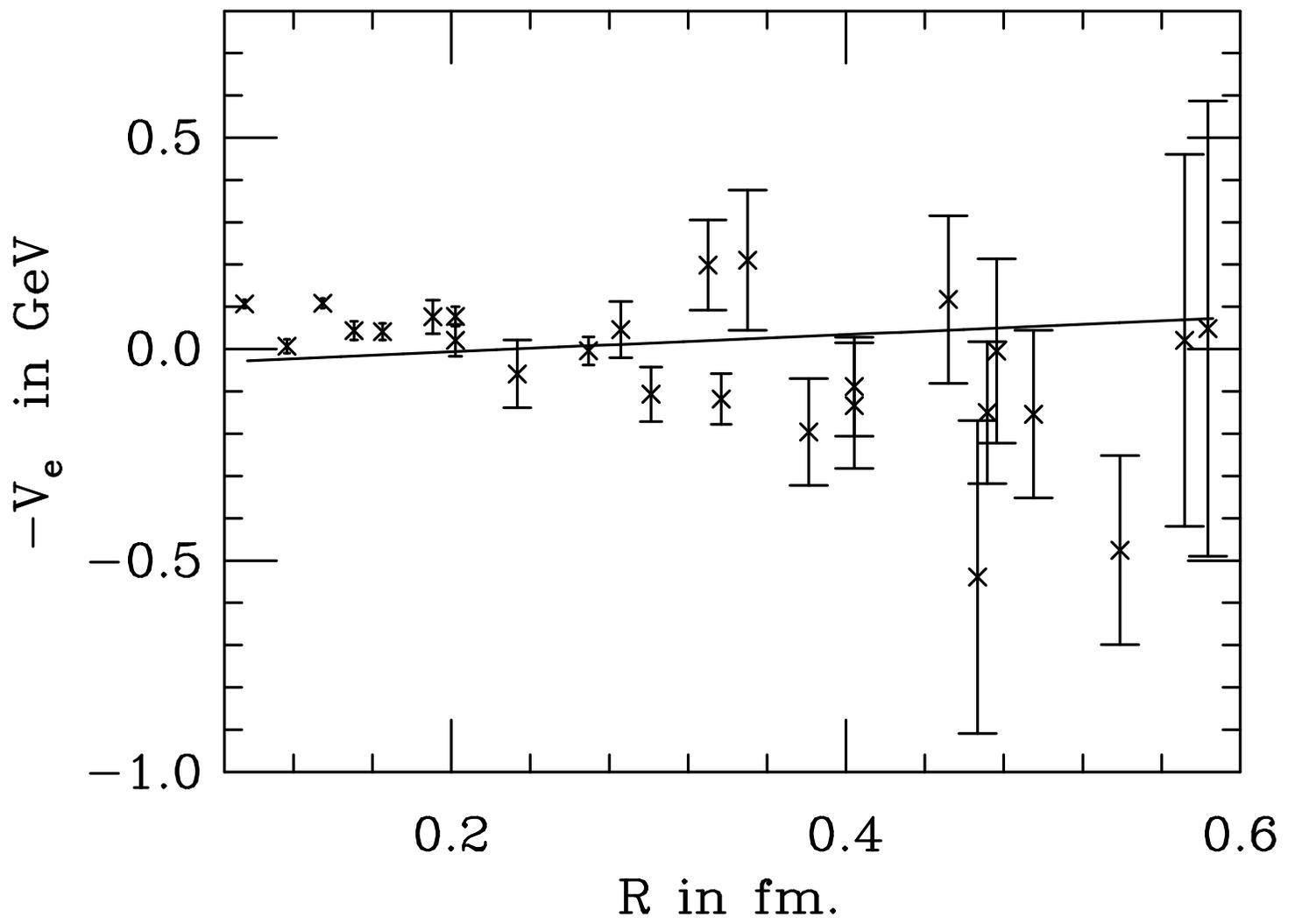
\end{document}